# Citation-based clustering of publications using CitNetExplorer and VOSviewer

Nees Jan van Eck and Ludo Waltman

Centre for Science and Technology Studies, Leiden University, The Netherlands
{ecknjpvan, waltmanlr}@cwts.leidenuniv.nl

Clustering scientific publications in an important problem in bibliometric research. We demonstrate how two software tools, CitNetExplorer and VOSviewer, can be used to cluster publications and to analyze the resulting clustering solutions. CitNetExplorer is used to cluster a large set of publications in the field of astronomy and astrophysics. The publications are clustered based on direct citation relations. CitNetExplorer and VOSviewer are used together to analyze the resulting clustering solutions. Both tools use visualizations to support the analysis of the clustering solutions, with CitNetExplorer focusing on the analysis at the level of individual publications and VOSviewer focusing on the analysis at an aggregate level. The demonstration provided in this paper shows how a clustering of publications can be created and analyzed using freely available software tools. Using the approach presented in this paper, bibliometricians are able to carry out sophisticated cluster analyses without the need to have a deep knowledge of clustering techniques and without requiring advanced computer skills.

## 1. Introduction

Clustering techniques play a prominent role in bibliometric research. They are for instance used to identify groups of related publications, authors, or journals. Clustering techniques have been developed mainly in fields such as statistics, computer science, and network science. Bibliometricians usually do not develop their own clustering techniques, but they use existing clustering techniques developed in other fields. They apply these techniques to bibliometric data sets, sometimes after adapting the techniques to the specific characteristics of bibliometric data.

When the number of objects to be clustered is relatively limited (e.g., at most a few hundred objects), analyzing and interpreting the results obtained from a clustering technique usually does not cause any significant difficulties. However, when dealing with large numbers of objects, analyzing and interpreting a clustering solution is far from straightforward. This can be a problem especially when clustering techniques are applied at the level of individual publications. We may then have clustering solutions that include many thousands or even many millions of publications (e.g., Boyack & Klavans, 2014; Klavans & Boyack, in press; Waltman & Van Eck, 2012). Making sense of these clustering solutions can be a serious challenge.

In this paper, our aim is to demonstrate how two software tools that we have developed, CitNetExplorer (Van Eck & Waltman, 2014a, 2014b; www.citnetexplorer.nl) and VOSviewer (Van Eck & Waltman, 2010, 2014b; www.vosviewer.com), can be used to cluster publications and to analyze the resulting clustering solutions. We use CitNetExplorer to cluster publications based on their citation relations and to analyze the resulting clustering solutions at the level of individual publications. We use VOSviewer to analyze the clustering solutions obtained using CitNetExplorer at an aggregate level. CitNetExplorer and VOSviewer both rely strongly on visualizations to facilitate the analysis of clustering solutions.

CitNetExplorer, which is an abbreviation of 'citation network explorer', is a software tools that we have developed for analyzing and visualizing citation networks. In the approach that we take in this paper, we first use CitNetExplorer to cluster publications based on their citation relations. For this purpose, CitNetExplorer employs a clustering technique that we have introduced in earlier papers (Waltman & Van Eck, 2012, 2013). We then use CitNetExplorer to analyze the resulting clustering solution at the level of individual publications. To facilitate the analysis of a clustering solution, the following features of CitNetExplorer are essential:

- *Visualizing a citation network.* CitNetExplorer can be used to visualize a citation network of publications, with publications shown along a time axis and with colors indicating the clusters to which publications belong. Using the visualization functionality of CitNetExplorer, we obtain an overview of the most frequently cited publications in a citation network, the citation relations between these publications, and the clusters to which the publications belong.
- *Drilling down into a citation network.* The drill down functionality of CitNetExplorer can be used to analyze a clustering solution at different levels of detail. We may for instance start with a visualization at the level of the entire citation network. We may then perform a drill down into one or more selected clusters, after which we are provided with a visualization at the level of the subnetwork consisting of the publications belonging to the selected clusters.

- *Searching for publications.* We can search for publications based on title, publication year, author name, and journal name. The search functionality of CitNetExplorer can be used to find publications that are of special interest, for instance all publications in a specific journal, and to find out to which clusters these publications belong.

VOSviewer is a software tool for constructing and visualizing bibliometric networks. In this paper, VOSviewer is used to complement CitNetExplorer. While we use CitNetExplorer to analyze a clustering solution at the level of individual publications, we use VOSviewer to analyze a clustering solution at an aggregate level. Two visualizations provided by VOSviewer play an important role. The first visualization shows the clusters in a clustering solution and the citation relations between these clusters. The second visualization uses a so-called term map to indicate the topics that are covered by a cluster. This visualization shows the most important terms occurring in the publications belonging to a cluster and the co-occurrence relations between these terms.

This paper is organized as follows. Section 2 discusses the clustering technique that is used by CitNetExplorer to cluster publications based on their citation relations. Section 3 demonstrates the use of CitNetExplorer and VOSviewer to cluster publications and to analyze the resulting clustering solutions. CitNetExplorer is used to cluster more than 100,000 publications in the field of astronomy and astrophysics, and CitNetExplorer and VOSviewer are used together to analyze the resulting clustering solutions. Section 4 concludes the paper.

## 2. Clustering technique

In this paper, we use the clustering technique that is available in the CitNetExplorer software tool. This section provides a discussion of this clustering technique. Subsection 2.1 explains how the relatedness of publications is determined, and Subsection 2.2 describes how publications are assigned to clusters. We refer to Waltman and Van Eck (2012, 2013) for a more extensive discussion of our clustering technique.

## 2.1. Determining the relatedness of publications

To cluster publications, we first need to determine the relatedness of publications. In the bibliometric literature, the most commonly used approaches to determine the relatedness of publications are based on either citation relations or word relations (for a more extensive discussion, see Van Eck & Waltman, 2014b). In the case of citation relations, a further distinction can be made between direct citation relations, bibliographic coupling relations, and co-citation relations (e.g., Boyack & Klavans, 2010; Klavans & Boyack, in press). In the case of word relations, shared words in the titles, abstracts, or full texts of publications serve as an indication of the relatedness of publications (e.g., Boyack et al., 2011; Janssens, Leta, Glänzel, & De Moor, 2006). Sometimes the relatedness of publications is determined using a combined approach that takes into account both citation relations and word relations (e.g., Boyack & Klavans, 2010; Janssens, Glänzel, & De Moor, 2008).

Our clustering technique determines the relatedness of publications based on direct citation relations. We prefer to use citation relations rather than word relations because the use of word relations involves some difficulties. Some words have a different meaning in different fields of science. These words may incorrectly indicate that publications from different fields are related to each other. Also, some words are very general and are used in many different fields. These words do not provide useful information on the relatedness of publications.

We prefer to use direct citation relations rather than bibliographic coupling relations (i.e., relations between publications that cite the same publication) or co-citation relations (i.e., relations between publications that are cited by the same publication) for two reasons. First, bibliographic coupling and co-citation relations are indirect relations, and we therefore expect them to provide less accurate information on the relatedness of publications than direct citation relations (Waltman & Van Eck, 2012). Second, there are many more bibliographic coupling or co-citation relations between publications than direct citation relations, and therefore the use of bibliographic coupling or co-citation relations may easily lead to computational problems. (This also applies to the use of word relations.) Although we prefer the use of direct citation relations over the use of bibliographic coupling or co-citation relations, we acknowledge that the use of direct citation relations also has a disadvantage. Within the period of analysis, some publications may have no direct citation relations with other publications. When using direct citation relations, these publications cannot be properly assigned to a cluster. This problem is especially serious when the period of analysis is relatively short. When using bibliographic coupling relations rather than direct citation relations, one usually does not have this problem. We note that, in addition to our own work, the use of direct citation relations is also advocated in recent work by Klavans and Boyack (in press).

**2.2. Clustering publications**

After the relatedness of publications has been determined, our clustering technique assigns publications to clusters. Each publication is assigned to exactly one cluster. Hence, there is no overlap of clusters and there are no publications without a cluster assignment. It may be argued that there should be room for publications to be assigned to more than one cluster. However, allowing publications to be assigned to multiple clusters introduces significant technical challenges. For this reason, we prefer to assign publications to a single cluster only. For most publications, we believe that it is reasonable to assign them to just one cluster.

Publications are assigned to clusters by maximizing a quality function. The quality function that is used has been introduced in an earlier paper (Waltman & Van Eck, 2012). This quality function is a variant of the well-known modularity function of Newman and Girvan (2004) and Newman (2004) developed in the field of network science. The quality function is very similar to the quality function resulting from the so-called constant Potts model proposed by Traag, Van Dooren, and Nesterov (2011). Our quality function has an important advantage over the popular modularity function. The modularity function suffers from a problem known as the resolution limit (Fortunato & Barthélemy, 2007). This problem causes the modularity function to yield counterintuitive results in certain situations. As shown by Traag et al. (2011), our quality function does not suffer from the resolution limit problem.

More specifically, our clustering technique assigns publications to clusters by maximizing the quality function

$$Q(x_1,\ldots,x_n) = \sum_{i=1}^{n}\sum_{j=1}^{n} \delta(x_i, x_j)\left(a_{ij} - \frac{\gamma}{2n}\right), \tag{1}$$

where *n* denotes the number of publications, $a_{ij}$ denotes the relatedness of publication *i* with publication *j*, $\gamma$ denotes a so-called resolution parameter, and $x_i$ denotes the cluster to which publication *i* is assigned. The function $\delta(x_i, x_j)$ equals 1 if $x_i = x_j$ and 0 otherwise. The relatedness of publication *i* with publication *j* is given by

$$a_{ij} = \frac{c_{ij}}{\sum_{k=1}^{n} c_{ik}}, \qquad (2)$$

where $c_{ij}$ equals 1 if either publication *i* cites publication *j* or publication *j* cites publication *i* and $c_{ij}$ equals 0 otherwise. Hence, if there is a direct citation relation between publications *i* and *j*, the relatedness of publication *i* with publication *j* is inversely proportional to the total number of direct citation relations of publication *i*. If there is no direct citation relation between publications *i* and *j*, the relatedness of the publications equals 0. Notice that our clustering technique ignores the direction of a citation (i.e., no distinction is made between publication *i* citing publication *j* and publication *j* citing publication *i*).

The value of the resolution parameter $\gamma$ in (1) should be chosen based on the purpose of the cluster analysis. Higher values of this parameter will yield a larger number of clusters. In other words, the higher the value of $\gamma$, the higher the level of detail of the clustering solution that will be obtained. In CitNetExplorer, the default value of $\gamma$ is 1. However, we emphasize that there is no generally optimal value of $\gamma$. Our recommendation to users of our clustering technique is to try out different values of $\gamma$ and to choose the value that seems to give the most useful results for the specific needs of a user.

In order to maximize the quality function in (1), our clustering technique uses the smart local moving algorithm introduced by Waltman and Van Eck (2013). This algorithm offers a more sophisticated alternative to the popular Louvain algorithm for modularity optimization (Blondel, Guillaume, Lambiotte, & Lefebvre, 2008). When the smart local moving algorithm and the Louvain algorithm are given a similar amount of computing time, the smart local moving algorithm typically identifies a clustering solution with a significantly higher value for the quality function. We refer to Waltman and Van Eck (2013) for an extensive comparison of the two algorithms.

Our clustering technique usually identifies a relatively limited number of larger clusters and a more substantial number of smaller clusters. Sometimes clusters are very small and for instance include only one or two publications. Because in many cases small clusters are of limited interest, a minimum cluster size parameter can be specified. Clusters that are too small can be either discarded or merged with other clusters. We refer to Waltman and Van Eck (2012) for a discussion of the approach that we take to merge small clusters with larger ones.

## 3. Results

We now demonstrate how CitNetExplorer and VOSviewer can be used to cluster publications and to analyze the resulting clustering solutions. In our demonstration, we work with a large data set of publications in the field of astronomy and astrophysics. We emphasize that in this paper it is not our aim to assess the quality of our clustering solutions or to compare our clustering solutions with other alternative solutions. We do not have the domain knowledge required to provide an in-depth interpretation of our clusters and to assess their quality. For a comparison of our clustering solutions with other alternative solutions, we refer to the comparison paper by Velden, Boyack, Gläser, Koopman, Scharnhorst, and Wang (in press) in this special issue.

### 3.1. Data

We use the 'Astro data set' that is also used in other papers in this special issue. A general introduction to the data set is provided in the introductory paper by Gläser, Glänzel, and Scharnhorst (in press) in this special issue. The data set was extracted from the Web of Science bibliographic database. It includes all publications of the document types 'article', 'letter', and 'proceedings paper' published between 2003 and 2010 in journals belonging to the Web of Science subject category 'Astronomy & Astrophysics'. The number of publications in the data set is 111,616. The publications appeared in 59 different journals. Of the 4,311,953 cited references provided in the publications in the data set, 929,364 point to publications in the data set. The statistics for the data set are summarized in Table 1.

CitNetExplorer requires a citation network to be acyclic. When analyzing a citation network, CitNetExplorer will make sure that the network is acyclic by removing citation relations that cause the network to have cycles. CitNetExplorer will also remove citation relations for which the citing publication appeared in an earlier year than the cited publication (e.g., a publication from 2009 citing a publication from 2010). In the case of our data set, of the 929,364 citation relations between publication in the data set, 3,824 were removed by CitNetExplorer. Hence, the citation network analyzed using CitNetExplorer included 925,540 citation relations.

Table 1. Statistics for the data set of astronomy and astrophysics publications.

| | |
|---|---:|
| No. of publications | 111,616 |
| No. of journals | 59 |
| No. of cited references | 4,311,953 |
| No. of citation relations between publications in the data set | 929,364 |
| No. of citation relations in CitNetExplorer | 925,540 |

**3.2. Using CitNetExplorer to cluster publications**

We clustered the publications in our data set using the clustering technique that is available in CitNetExplorer. We refer to Section 2 for a discussion of this clustering technique. Our citation network of 111,616 publications has a largest component that includes 101,828 publications. Only these 101,828 publications were included in the cluster analysis. The other 9,788 publications were not assigned to a cluster.

As already explained in Subsection 2.2, clustering solutions can be created at different levels of detail. The choice of the most suitable level of detail is not a technical one but instead depends on the purpose of the cluster analysis. Our recommendation is to create multiple clustering solutions at different levels of detail and to use the solution (or the solutions) that fits best with the needs one has. In line with this idea, we used CitNetExplorer to create four clustering solutions, each providing a different level of detail. The clustering solutions are based on different values of the resolution parameter and the minimum cluster size parameter. Clusters that did not meet the minimum cluster size criterion were merged with larger clusters. We note that the four clustering solutions do not have a hierarchical relationship with each other. For instance, a cluster in the most detailed clustering solution may overlap with more than one cluster in the second most detailed clustering solution.

For each of the four clustering solutions, Table 2 reports the values of the resolution parameter and the minimum cluster size parameter. The table also provides for each clustering solution a number of statistics. These are the number of clusters, the average number of publications per cluster, and the number of publications in the smallest and the largest cluster. As can be seen in Table 2, the clustering solution that provides the lowest level of detail, referred to as the level 1 clustering, includes 22 clusters. This clustering solution has an average cluster size of 4,629 publications and a maximum cluster size of almost 15,000 publications. On the other hand, the clustering solution that provides the highest level of detail, referred to as the level 4 clustering, includes 434 clusters. This clustering solution has an average cluster size of 235 publications and a maximum cluster size of somewhat more than 1,000 publications. The statistics reported in Table 2 make clear that, regardless of the level of detail of a clustering solution, the distribution of publications over clusters is quite skewed. This is a typical phenomenon when our technique for clustering publications is used (for more details, see Waltman & Van Eck, 2012).

Table 2. Parameters and statistics for the different clustering solutions.

| Level | Resolution | Min. cluster size | No. of clusters | Avg. no. of pub. per cluster | No. of pub. smallest cluster | No. of pub. largest cluster |
| --- | --- | --- | --- | --- | --- | --- |
| 1 | 1.8 | 500 | 22 | 4,628.5 | 794 | 14,873 |
| 2 | 3.0 | 250 | 42 | 2,424.5 | 253 | 9,395 |
| 3 | 10.0 | 150 | 115 | 885.5 | 176 | 2,891 |
| 4 | 40.0 | 50 | 434 | 234.6 | 50 | 1,080 |

In the rest of this paper, our focus will be mainly on the level 1 clustering. To get an impression of the topics covered by the 22 level 1 clusters, Table 3 presents for each cluster the number of publications and five characteristic terms. The characteristic terms were extracted from the titles of the publications belonging to a cluster using the methodology described by Waltman and Van Eck (2012). A more extensive summary of the level 1 clusters is provided in Table A1 in the appendix. For each cluster, this table lists not only the number of publications and five characteristic terms but also the three journals with the largest number of publications and the most frequently cited publication. In addition, for each cluster, ten standardized terms are presented. These terms were selected using a standardized approach that has also been used in other papers in this special issue.

Table 3. Brief summary of the 22 level 1 clusters.

| Cluster | No. of pub. | Terms |
|---|---|---|
| 1 | 14,873 | galaxy cluster; galaxy; early type galaxy; abell; high redshift |
| 2 | 8,954 | dark energy; inflation; wmap; cosmic microwave background; cosmology |
| 3 | 7,998 | solar flare; coronal mass ejection; solar corona; solar cycle; sunspot |
| 4 | 7,483 | brown dwarf; protoplanetary disk; extrasolar planet; planet; exoplanet |
| 5 | 5,704 | molecular cloud; region; dark cloud; protostar; dense core |
| 6 | 5,597 | globular cluster; globular cluster system; metal poor star; star cluster; omega centauri |
| 7 | 5,363 | qcd; lattice; decay; finite temperature; lattice qcd |
| 8 | 5,211 | cern lhc; lhc; dark matter annihilation; leptogenesis; higgs boson |
| 9 | 5,179 | ultraluminous x ray source; cygnus x; x ray binary; microquasar; integral |
| 10 | 3,904 | quasinormal mode; hawking radiation; ads; higher dimension; wormhole |
| 11 | 3,527 | supernova remnant; pulsar; psr; magnetar; radio pulsar |
| 12 | 3,413 | asteroid; comet; body problem; trans neptunian object; centaur |
| 13 | 3,392 | eta carinae; asteroseismology; ap star; peculiar star; binary |
| 14 | 3,355 | titan; mars; venus; mercury; europa |
| 15 | 3,182 | grb; gamma ray burst; afterglow; type ia supernovae; short gamma ray burst |
| 16 | 3,156 | lisa; gravitational wafe; numerical relativity; gravitational wave detector; gravitational wave burst |
| 17 | 2,625 | blazar; ultra high energy cosmic ray; bl lacertae object; pks; bl lac object |
| 18 | 2,228 | iri; cluster observation; ionosphere; magnetosheath; low latitude |
| 19 | 2,088 | cataclysmic variable; white dwarf; nova; superoutburst; dwarf novae |
| 20 | 1,963 | loop quantum gravity; loop quantum cosmology; quantum gravity; lorentz; lorentz violation |
| 21 | 1,839 | planetary nebulae; symbiotic star; planetary nebula ngc; central star; planetary nebula |
| 22 | 794 | pioneer; lense thirring effect; teleparallel gravity; equivalence principle; iau |

**3.3. Using CitNetExplorer to analyze clustering solutions at the publication level**

We first use CitNetExplorer to analyze the level 1 clustering. The analysis takes place at the level of individual publications. In the next subsection, we use VOSviewer to perform an analysis at an aggregate level.

For a given set of publications, CitNetExplorer can be used to get an overview of the most frequently cited publications, the citation relations between these publications, the temporal order of the publications, and the assignment of the publications to clusters. Suppose we are interested to get a better understanding of the publications belonging to level 1 clusters 1, 2, 3, and 4 (i.e., the four largest level 1 clusters). Figure 1 provides a CitNetExplorer visualization of the 100 most frequently cited publications in these four clusters. Each publication is indicated by a circle, and publications are labeled by the last name of the first author. The vertical dimension represents time, with publications in the top part of the visualization being older and publications in the bottom part being more recent. In the horizontal dimension, publications are positioned based on their relatedness in terms of citations. Publications that are strongly related in terms of citations, taking into account not only direct citation relations between publications but also indirect citation relations, tend to be located close to each other in the horizontal dimension. Publications that are only weakly related in terms of citations are located further away from each other. The curved lines between publications indicate citation relations, with the citing publication always being located below the cited publication. The darker lines represent direct citation relations, while the lighter lines represent indirect citation relations. There is an indirect citation relation from publication A to publication B if publication A does not directly cite publication B but if publication A for instance cites publication C and publication C in turn cites publication B. The color of a publication indicates the cluster to which the publication belongs, with blue, green, purple, and orange corresponding with, respectively, clusters 1, 2, 3, and 4.

The visualization provided in Figure 1 is static. In the CitNetExplorer software tool, the same visualization is presented in an interactive way. This for instance makes it possible to zoom in on a specific area in the visualization and to explore in more detail the publications located in that area. Also, by hovering the mouse over a publication, bibliographic information on the publication is presented, for instance the authors, the title, and the journal in which the publication appeared.

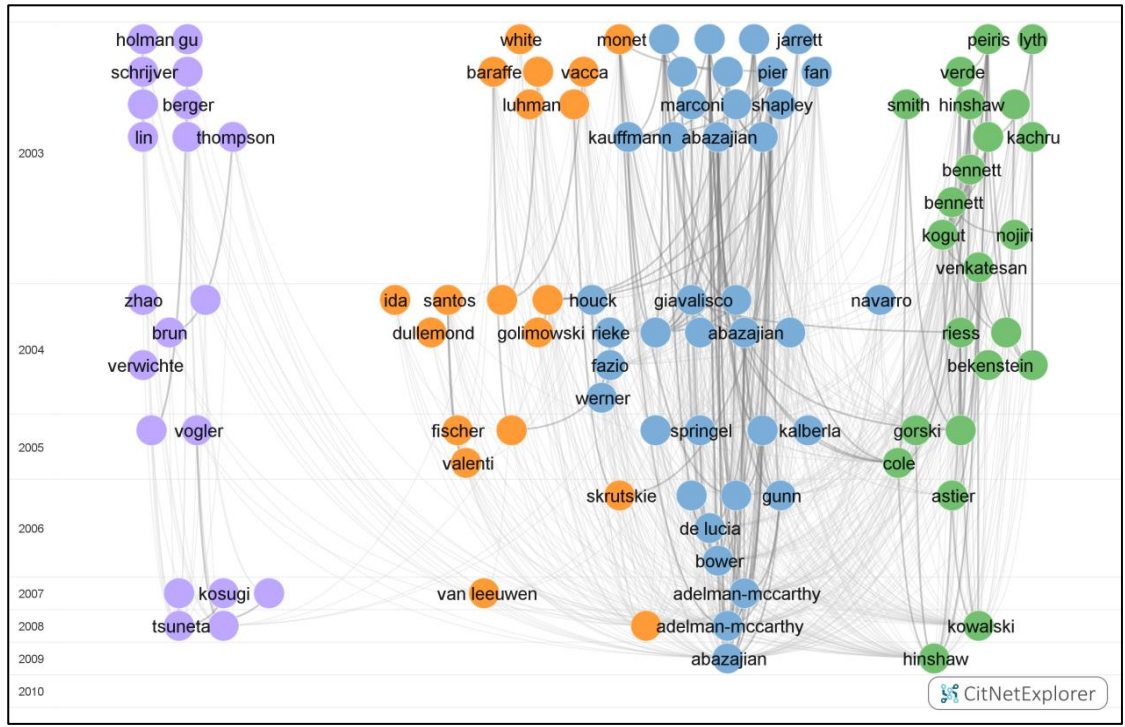

Figure 1. CitNetExplorer visualization of the 100 most frequently cited publications in level 1 clusters 1, 2, 3, and 4. Colors indicate the level 1 cluster to which a publication belongs.

What do we learn from the visualization provided in Figure 1? First of all, the visualization confirms that level 1 clusters 1, 2, 3, and 4 cover relatively independent bodies of literature. Most citation relations shown in the visualization are between publications belonging to the same cluster rather than between publications belonging to different clusters. In addition, the visualization reveals that clusters 1, 2, and 4 (shown in blue, green, and orange, respectively) are more strongly connected to each other than to cluster 3 (shown in purple), at least when focusing on the most highly cited publications in the different clusters. Of the four clusters, cluster 3 therefore appears to be the one that is most independent from the others.

A more detailed interpretation of the visualization presented in Figure 1 requires expert knowledge of the field of astronomy and astrophysics. Using the visualization, an expert in the field obtains a basic understanding of the topics covered by the different clusters and of the developments taking place within each cluster. An expert will probably be familiar with many of the publications shown in the visualization and will have some general idea of the role played by these publications in the development of the field of astronomy and astrophysics. By combining this expert knowledge with the information offered by the visualization, on the one hand an expert can provide an interpretation of the clusters and on the other hand the expert can deepen his or her understanding of the astronomy and astrophysics field.

Suppose next that we would like to explore level 1 cluster 2 in more detail. This can be done using the drill down functionality of CitNetExplorer. This functionality makes it possible to drill down into a specific subnetwork of a citation network. In this case, a drill down is performed into the subnetwork consisting of the publications belonging to cluster 2 and the citation relations between these publications. After drilling down, the visualization presented in Figure 2 is obtained. Of the 8,954 publications belonging to cluster 2, the visualization shows the 100 most frequently cited ones. As discussed in Subsection 3.2, publications were clustered at four levels of detail. In the visualization, the color of a publication is determined by the cluster to which the publication belongs in the level 3 clustering. As can be seen in the visualization, the most frequently cited publications in level 1 cluster 2 belong mostly to three different level 3 clusters. These clusters are indicated using the colors red, brown, and light blue in the visualization.

The visualization presented in Figure 2 provides insight into the subdivision of level 1 cluster 2 into smaller level 3 clusters. If a deeper understanding of the literature is required, one could perform a further drill down. In this way, a specific level 3 cluster could be explored in more detail. In a next step, another drill down could be performed to explore an even smaller level 4 cluster.

An analysis using CitNetExplorer takes place at the level of individual publications. In many cases, one may also want to analyze a clustering solution at an aggregate level. This is not possible using CitNetExplorer, but it can be accomplished using other software tools. In particular, VOSviewer can be used for this purpose, as discussed in the next subsection.

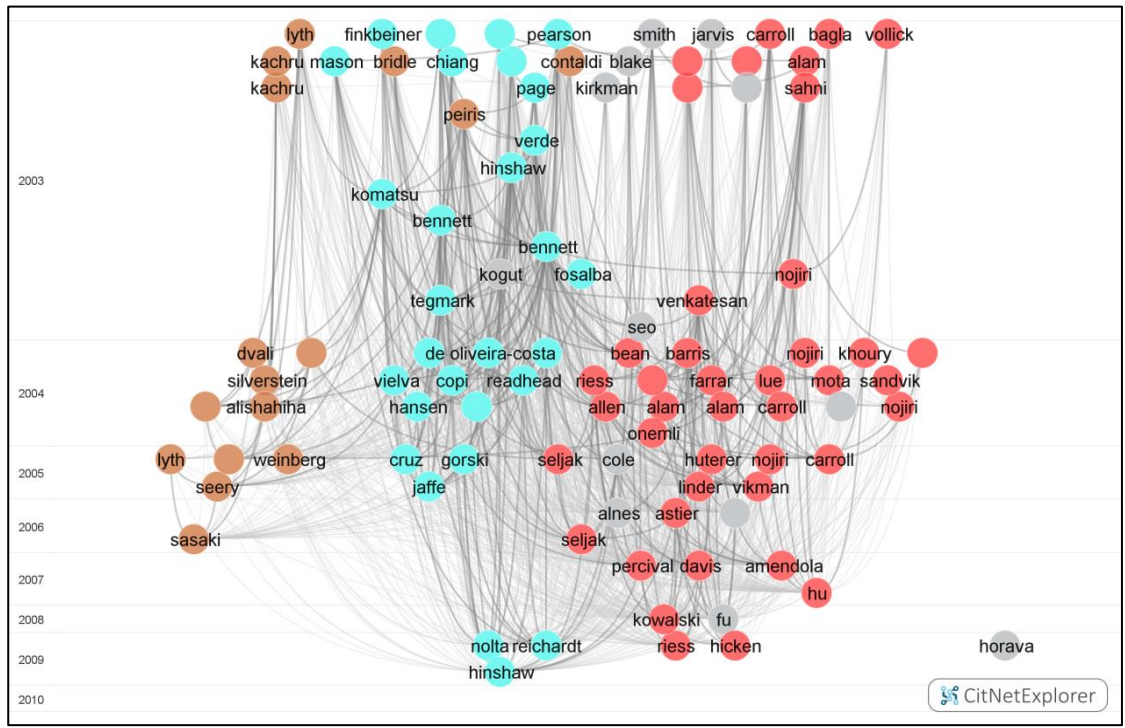

Figure 2. CitNetExplorer visualization of the 100 most frequently cited publications in level 1 cluster 2. Colors indicate the level 3 cluster to which a publication belongs.

**3.4. Using VOSviewer to analyze clustering solutions at an aggregate level**

We now use VOSviewer to carry out a further analysis of the level 1 clustering. The analysis is performed at an aggregate level and uses two visualizations. One visualization shows the level 1 clusters and the citation relations between these clusters. The other visualization uses a term map to indicate the topics that are covered by a level 1 cluster.

A visualization of the 22 level 1 clusters and their citation relations is provided in Figure 3. In this visualization, the size of a cluster reflects the number of publications belonging to the cluster. Larger clusters include more publications. The distance between two clusters approximately indicates the relatedness of the clusters in terms of citations. Clusters that are located close to each other tend to be strongly related in terms of citations, while clusters that are located further away from each other tend to be less strongly related. The curved lines between the clusters also reflect the relatedness of clusters, with the thickness of a line representing the number of citations between two clusters. The horizontal and vertical axes have no special meaning.

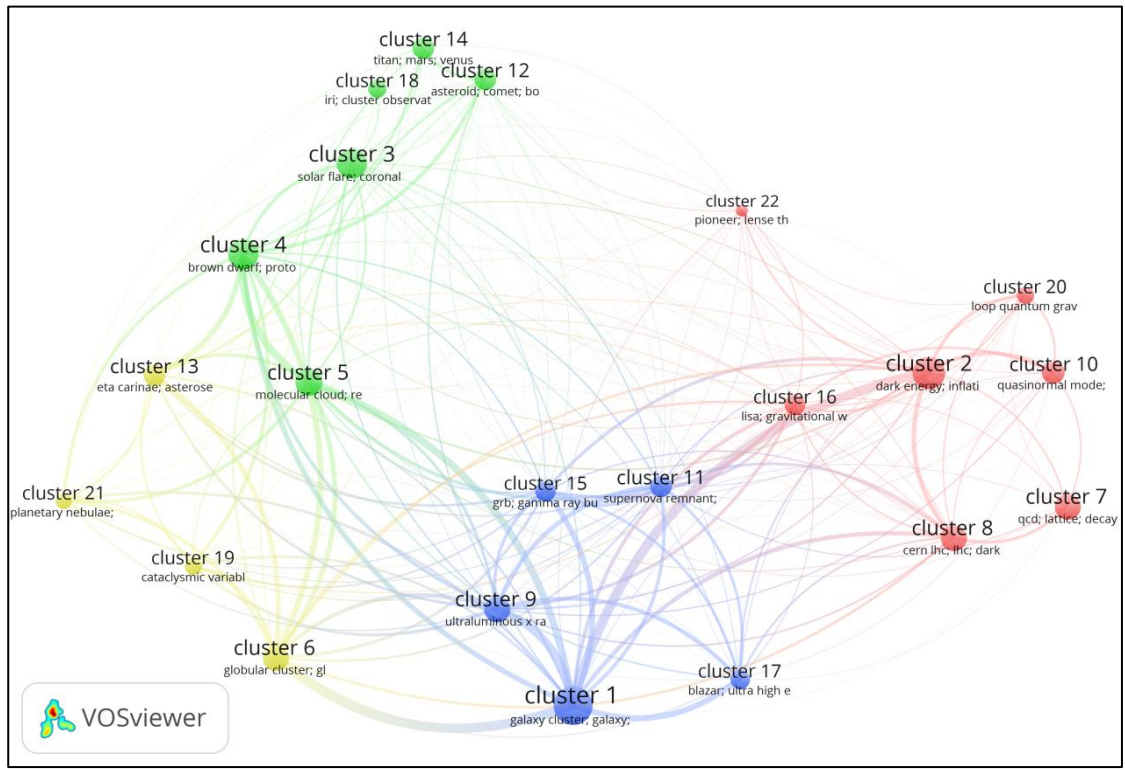

Figure 3. VOSviewer visualization of the 22 level 1 clusters and their citation relations. An interactive version of the visualization is available online at http://goo.gl/968hLw.

VOSviewer has its own clustering technique (Waltman, Van Eck, & Noyons, 2010), and this clustering technique was used to partition the 22 level 1 clusters into four groups. This was done based on the citation relations between the clusters. In the visualization presented in Figure 3, each cluster has a color (i.e., red, green, blue, or yellow) that indicates the group to which the cluster was assigned. In this way, a breakdown of the astronomy and astrophysics literature into broad subfields is obtained. A rough interpretation of the visualization is as follows. The red clusters in the right area of the visualization seem to cover research in astroparticle physics, gravitational physics, and cosmology. The blue and yellow clusters in the bottom area seem to cover astrophysics research on galaxies and stars. The clusters in the top-left area, colored green, seem to relate to research in solar physics and planetary science.

An interactive version of the visualization provided in Figure 3 is available online at http://goo.gl/968hLw. The interactive visualization offers additional information not visible in Figure 3. In particular, when the mouse is hovered over a cluster, more detailed information on the cluster is presented, similar to the information provided in Table 3.

Suppose now that we would like to get a better understanding of a specific level 1 cluster, for instance cluster 3. For this purpose, we use the term map visualization presented in Figure 4. To create this visualization, the titles and abstracts of the 7,998 publications belonging to cluster 3 were analyzed using natural language processing techniques (Van Eck & Waltman, 2011). For each publication, the terms occurring in the title and abstract of the publication were identified. Of all terms that were found in at least 15 publications, the 1,420 terms that seemed most relevant were algorithmically selected. These terms are shown in the term map visualization provided in Figure 4. Each term is represented by a circle, and some terms are also indicated by a label. (VOSviewer aims to avoid overlapping labels, and therefore labels are visible only for some of the terms.) The size of a term reflects the number of publications in which the term was found, and the distance between two terms offers an approximate indication of the relatedness of the terms. The relatedness of terms was determined based on co-occurrences. In other words, the larger the number of publications in which two terms were both found, the stronger the relation between the terms. Colors represent groups of terms that are relatively strongly related to each other. These groups were identified using the clustering technique of VOSviewer that was also mentioned above. In the visualization, the strongest relations between terms are also indicated using curved lines.

What does the term map visualization tell us about the topics that are covered by level 1 cluster 3? Publications belonging to cluster 3 seem to study various types of solar phenomena. In the right area of the visualization, we observe terms dealing with the phenomenon of solar wind and the related phenomenon of coronal mass ejection. In the top area, terms related to the phenomenon of solar flares can be found. Terms related to the phenomenon of sunspots are located in the left area of the visualization. In the bottom area, we observe the term 'solar cycle'. Solar phenomena are often influenced by the solar cycle.

An interactive version of the term map visualization presented in Figure 4 is available online at http://goo.gl/sotbF1. In the interactive visualization, it is possible to zoom in on specific areas in the visualization. When zooming in, the labels of more and more terms become visible, making it possible to interpret a specific area in the visualization in more detail.

Figure 4. VOSviewer term map visualization for level 1 cluster 3. The visualization shows 1,420 terms extracted from the titles and abstracts of the publications belonging to the cluster. The strongest co-occurrence relations between terms are shown as well. An interactive version of the visualization is available online at http://goo.gl/sotbF1.

## 4. Conclusion

We have demonstrated the use of CitNetExplorer and VOSviewer for clustering publications based on direct citation relations and for analyzing the resulting clustering solutions. We have shown how the visualizations provided by the two software tools complement each other, with CitNetExplorer focusing on visualizations at the level of individual publications and VOSviewer focusing on visualizations at an aggregate level.

Bibliometricians usually do not develop their own clustering techniques, but instead they apply existing clustering techniques in a bibliometric context. The approach presented in this paper is well suited for this purpose. The software tools that we have used are freely available. Using these tools, publications can be clustered without the need to have a deep knowledge of clustering techniques. In addition, no advanced computer skills are required. For instance, data download from the online Web of Science database can be provided directly as input to the software tools, without the need to preprocess the data. Of course, despite the ease of use of our tools, a basic understanding of clustering techniques remains essential to perform meaningful analyses and to avoid misinterpretations of the results that are obtained.

The clustering technique that we have used is based on recent developments in the fields of network science and bibliometrics (Traag et al., 2011; Waltman & Van Eck, 2012, 2013). In addition to our own work, this clustering technique has also been used in the work of other bibliometricians (Boyack & Klavans, 2014; Klavans & Boyack, in press; Small, Boyack, & Klavans, 2014). Our clustering technique determines the relatedness of publications based on direct citation relations. A major advantage of the use of direct citation relations is the possibility to efficiently cluster very large numbers of publications (e.g., tens of millions of publications). A disadvantage is that, due to a lack of direct citation relations, some publications cannot be properly assigned to a cluster. We note that, in addition to our clustering technique, other clustering techniques could also be considered for clustering publications based on direct citation relations. For instance, in a recent study (Šubelj, Van Eck, & Waltman, 2015), we found indications suggesting that the map equation technique, used together with the Infomap optimization algorithm (Bohlin, Edler, Lancichinetti, & Rosvall, 2014; Rosvall & Bergstrom, 2008), may give particularly good results.

We have demonstrated the capabilities of CitNetExplorer and VOSviewer for clustering publications and for analyzing the resulting clustering solutions. However, the combined use of the two software tools is somewhat laborious, and preparing the input data for VOSviewer based on the clustering results provided by CitNetExplorer is not entirely straightforward. In future research, we therefore plan to work on the development of a single integrated software tool in which many of the key features of CitNetExplorer and VOSviewer are brought together. We have in mind a tool that combines different types of interactive visualizations to support users in exploring the scientific literature. A technique for clustering publications based on direct citation relations, similar to the technique used in this paper, will be at the core of the new tool. Like in this paper, it will be possible to create clustering solutions at different levels of detail. The new tool will provide interactive functionality for browsing through a hierarchical structure of clusters, and the tool will use visualizations similar to the ones used in this paper to show citation relations between publications and between clusters and to indicate the topics covered by clusters. The dynamics of clusters, revealing for instance how interest in a topic has grown or declined over time, will be made visible as well.

## Acknowledgments

We would like to thank Julia Heuritsch for her help in the interpretation of the visualizations presented in Figures 3 and 4. We are grateful to Kevin Boyack, Jochen Gläser, Andrea Scharnhorst, and an anonymous referee for their comments on earlier drafts of this paper.

## Appendix

Table A1. Extended summary of the 22 level 1 clusters.

| Cluster | No. of pub. | Terms, standardized terms, journals, and most frequently cited publication |
|---|---|---|
| 1 | 14,873 (14.6%) | *Terms*: galaxy cluster; galaxy; early type galaxy; abell; high redshift |
| | | *Standardized terms*: galaxies, redshift, star formation, sample, active galactic, agn, gas, galaxy clusters, digital sky, sloan digital |
| | | *Journals*: astrophysical journal; monthly notices of the royal astronomical society; astronomy & astrophysics |
| | | *Publication*: fazio, gg; et al. (2004). the infrared array camera (irac) for the spitzer space telescope. astrophys j suppl s, 154(1), 10–17. |
| 2 | 8,954 (8.8%) | *Terms*: dark energy; inflation; wmap; cosmic microwave background; cosmology |
| | | *Standardized terms*: dark energy, microwave background, cosmic microwave, inflation, cosmological, universe, cmb, power spectrum, background cmb, scalar field |
| | | *Journals*: physical review d; journal of cosmology and astroparticle physics; monthly notices of the royal astronomical society |
| | | *Publication*: bennett, cl; et al. (2003). first-year wilkinson microwave anisotropy probe (wmap) observations: preliminary maps and basic results. astrophys j suppl s, 148(1), 1–27. |
| 3 | 7,998 (7.9%) | *Terms*: solar flare; coronal mass ejection; solar corona; solar cycle; sunspot |
| | | *Standardized terms*: solar, coronal, active region, cme, flare, magnetic field, sunspot, mass ejections, quiet sun, chromosphere |
|   |   | *Journals*: astrophysical journal; astronomy & astrophysics; solar physics |
|   |   | *Publication*: kosugi, t; et al. (2007). the hinode (solar-b) mission: an overview. sol phys, 243(1), 3–17. |
| 4 | 7,483 (7.3%) | *Terms*: brown dwarf; protoplanetary disk; extrasolar planet; planet; exoplanet |
|   |   | *Standardized terms*: planets, brown dwarfs, planet formation, transit, extrasolar planets, star, tauri stars, jup, giant planet, hd |
|   |   | *Journals*: astrophysical journal; astronomy & astrophysics; monthly notices of the royal astronomical society |
|   |   | *Publication*: skrutskie, mf; et al. (2006). the two micron all sky survey (2mass). astron j, 131(2), 1163–1183. |
| 5 | 5,704 (5.6%) | *Terms*: molecular cloud; region; dark cloud; protostar; dense core |
|   |   | *Standardized terms*: molecular cloud, protostellar, cloud, interstellar, young stellar, star forming, molecules, massive star, forming region, stellar objects |
|   |   | *Journals*: astrophysical journal; astronomy & astrophysics; monthly notices of the royal astronomical society |
|   |   | *Publication*: lada, cj; et al. (2003). embedded clusters in molecular clouds. annu rev astron astr, 41, 57–115. |
| 6 | 5,597 (5.5%) | *Terms*: globular cluster; globular cluster system; metal poor star; star cluster; omega centauri |
|   |   | *Standardized terms*: globular clusters, fe h, metal poor, giant branch, stars, red giant, metallicity, galactic globular, horizontal branch, milky way |
|   |   | *Journals*: astrophysical journal; astronomy & astrophysics; monthly notices of the royal astronomical society |
|   |   | *Publication*: zacharias, n; et al. (2004). the second us naval observatory ccd astrograph catalog (ucac2). astron j, 127(5), 3043–3059. |
| 7 | 5,363 (5.3%) | *Terms*: qcd; lattice; decay; finite temperature; lattice qcd |
|   |   | *Standardized terms*: qcd, quark, meson, lattice, decays, chiral, pi pi, gluon, pion, j psi |
|   |   | *Journals*: physical review d; international journal of modern physics d; classical and quantum gravity |
|   |   | *Publication*: ball, p; et al. (2005). new results on b ->pi,k,eta decay form factors from light-cone sum rules. phys rev d, 71(1), 014015. |
| 8 | 5,211 (5.1%) | *Terms*: cern lhc; lhc; dark matter annihilation; leptogenesis; higgs boson |
|   |   | *Standardized terms*: standard model, neutrino, higgs, lhc, minimal supersymmetric, lepton, supersymmetric standard, gev, top quark, hadron collider |
|   |   | *Journals*: physical review d; journal of cosmology and astroparticle physics; astroparticle physics |
|   |   | *Publication*: arkani-hamed, n; et al. (2009). a theory of dark matter. phys rev d, |

79(1), 015014.

| | | |
|---|---|---|
| 9 | 5,179 (5.1%) | *Terms*: ultraluminous x ray source; cygnus x; x ray binary; microquasar; integral |
| | | *Standardized terms*: x ray, ray binary, black hole, accretion disk, hard state, ray timing, neutron star, rossi x, timing explorer, xmm newton |
| | | *Journals*: astrophysical journal; astronomy & astrophysics; monthly notices of the royal astronomical society |
| | | *Publication*: winkler, c; et al. (2003). the integral mission. astron astrophys, 411(1), l1–l6. |
| 10 | 3,904 (3.8%) | *Terms*: quasinormal mode; hawking radiation; ads; higher dimension; wormhole |
| | | *Standardized terms*: black holes, spacetimes, horizon, ads, solutions, metric, dimensional, supergravity, static, spherically symmetric |
| | | *Journals*: physical review d; classical and quantum gravity; general relativity and gravitation |
| | | *Publication*: gauntlett, jp; et al. (2003). all supersymmetric solutions of minimal supergravity in five dimensions. classical quant grav, 20(21), 4587–4634. |
| 11 | 3,527 (3.5%) | *Terms*: supernova remnant; pulsar; psr; magnetar; radio pulsar |
| | | *Standardized terms*: pulsar, supernova remnant, psr, snr, neutron star, wind nebula, anomalous x, remnant snr, radio pulsars, magnetar |
| | | *Journals*: astrophysical journal; astronomy & astrophysics; monthly notices of the royal astronomical society |
| | | *Publication*: manchester, rn; et al. (2005). the australia telescope national facility pulsar catalogue. astron j, 129(4), 1993–2006. |
| 12 | 3,413 (3.4%) | *Terms*: asteroid; comet; body problem; trans neptunian object; centaur |
| | | *Standardized terms*: asteroid, comet, main belt, kuiper belt, meteor, perihelion, bodies, solar system, albedo, trans neptunian |
| | | *Journals*: icarus; astronomy & astrophysics; celestial mechanics & dynamical astronomy |
| | | *Publication*: bernstein, gm; et al. (2004). the size distribution of trans-neptunian bodies. astron j, 128(3), 1364–1390. |
| 13 | 3,392 (3.3%) | *Terms*: eta carinae; asteroseismology; ap star; peculiar star; binary |
| | | *Standardized terms*: eclipsing binary, star, asteroseismic, chemically peculiar, wilson devinney, delta scuti, eta carinae, contact binary, hd, pulsation |
| | | *Journals*: astronomy & astrophysics; monthly notices of the royal astronomical society; astrophysical journal |
| | | *Publication*: asplund, m; et al. (2004). line formation in solar granulation - iv. [o i], oi and oh lines and the photospheric o abundance. astron astrophys, 417(2), 751–768. |
| 14 | 3,355 | *Terms*: titan; mars; venus; mercury; Europa |
|    | (3.3%)       | *Standardized terms*: mars, titan, atmosphere, water, deposits, cassini, mars express, ice, venus, moon                                                                   |
|    |              | *Journals*: icarus; planetary and space science; advances in space research                                                                                              |
|    |              | *Publication*: smith, md (2004). interannual variability in tes atmospheric observations of mars during 1999-2003. icarus, 167(1), 148–165.                              |
| 15 | 3,182        | *Terms*: grb; gamma ray burst; afterglow; type ia supernovae; short gamma ray burst                                                                                      |
|    | (3.1%)       | *Standardized terms*: grb, ray bursts, gamma ray, afterglow, bursts grbs, sn, explosion, swift, type ia, supernova sn                                                    |
|    |              | *Journals*: astrophysical journal; monthly notices of the royal astronomical society; astronomy & astrophysics                                                           |
|    |              | *Publication*: gehrels, n; et al. (2004). the swift gamma-ray burst mission. astrophys j, 611(2), 1005–1020.                                                             |
| 16 | 3,156        | *Terms*: lisa; gravitational wafe; numerical relativity; gravitational wave detector; gravitational wave burst                                                           |
|    | (3.1%)       | *Standardized terms*: gravitational wave, lisa, inspiral, ligo, wave detectors, laser interferometer, binary black, waveforms, numerical relativity, post newtonian      |
|    |              | *Journals*: physical review d; classical and quantum gravity; astrophysical journal                                                                                      |
|    |              | *Publication*: heger, a; et al. (2005). presupernova evolution of differentially rotating massive stars including magnetic fields. astrophys j, 626(1), 350–363.         |
| 17 | 2,625        | *Terms*: blazar; ultra high energy cosmic ray; bl lacertae object; pks; bl lac object                                                                                    |
|    | (2.6%)       | *Standardized terms*: blazar, bl lac, jet, lac objects, radio galaxies, synchrotron, ultra high, radio, radio sources, 3c                                                |
|    |              | *Journals*: astrophysical journal; astronomy & astrophysics; monthly notices of the royal astronomical society                                                           |
|    |              | *Publication*: aharonian, f; et al. (2007). an exceptional very high energy gamma-ray flare of pks 2155-304. astrophys j, 664(2), l71–l74.                               |
| 18 | 2,228        | *Terms*: iri; cluster observation; ionosphere; magnetosheath; low latitude                                                                                                |
|    | (2.2%)       | *Standardized terms*: ionospheric, auroral, radar, substorm, geomagnetic, magnetopause, iri, tec, midnight, field aligned                                                |
|    |              | *Journals*: annales geophysicae; advances in space research; planetary and space science                                                                                 |
|    |              | *Publication*: milan, se; et al. (2003). variations in the polar cap area during two substorm cycles. ann geophys-germany, 21(5), 1121–1140.                             |
| 19 | 2,088        | *Terms*: cataclysmic variable; white dwarf; nova; superoutburst; dwarf novae                                                                                              |
|    | (2.1%)       | *Standardized terms*: white dwarf, nova, cataclysmic variable, dwarf nova, subdwarf b, wd, sdb, orbital period, sdb stars, superhumps                                    |
|    |              | *Journals*: astronomy & astrophysics; astrophysical journal; monthly notices of the royal astronomical society                                                           |
|    |        | *Publication*: ritter, h; et al. (2003). catalogue of cataclysmic binaries, low-mass x-ray binaries and related objects (seventh edition). astron astrophys, 404(1), 301–303. |
| 20 | 1,963 (1.9%) | *Terms*: loop quantum gravity; loop quantum cosmology; quantum gravity; lorentz; lorentz violation |
|    |        | *Standardized terms*: quantum gravity, loop quantum, noncommutative, quantum, quantization, quantum cosmology, algebra, spin foam, casimir, hilbert space |
|    |        | *Journals*: physical review d; classical and quantum gravity; international journal of modern physics d |
|    |        | *Publication*: ashtekar, a; et al. (2004). multipole moments of isolated horizons. classical quant grav, 21(11), 2549–2570. |
| 21 | 1,839 (1.8%) | *Terms*: planetary nebulae; symbiotic star; planetary nebula ngc; central star; planetary nebula |
|    |        | *Standardized terms*: planetary nebulae, pne, post agb, central star, asymptotic giant, nebulae pne, symbiotic, pn, mira, agb stars |
|    |        | *Journals*: astronomy & astrophysics; astrophysical journal; monthly notices of the royal astronomical society |
|    |        | *Publication*: van winckel, h (2003). post-agb stars. annu rev astron astr, 41, 391–427. |
| 22 | 794 (0.8%) | *Terms*: pioneer; lense thirring effect; teleparallel gravity; equivalence principle; iau |
|    |        | *Standardized terms*: teleparallel, nutation, lense thirring, lageos, laser ranging, gravitomagnetic, celestial reference, pioneer, gravitational field, grace |
|    |        | *Journals*: international journal of modern physics d; advances in space research; astronomy & astrophysics |
|    |        | *Publication*: pitjeva, ev (2005). high-precision ephemerides of planets - epm and determination of some astronomical constants. solar syst res+, 39(3), 176–186. |